%% file: ajj.tex
\documentclass[11pt]{article}

\include{setup}

\begin{document}

\thispagestyle{empty}
\def\thefootnote{\fnsymbol{footnote}}
\setcounter{footnote}{1}
\null
\vskip 0cm
\vfill
\begin{center}
 {\Large \boldmath{\bf Next-to-leading order QCD corrections to \\
 photon production via weak-boson fusion}
\par} \vskip 2.5em
{\large
{\sc B.\ J\"ager
}\\[2ex]
{\normalsize \it 
\it Institut f\"ur Theoretische Physik und Astrophysik, Universit\"at W\"urzburg, 97074 W\"urzburg, Germany
}\\[1ex]
}
\par \vskip 5em
\end{center}\par
\vskip .0cm \vfill {\bf Abstract:} \par
We present a calculation of next-to-leading order QCD corrections to $\ajj$ production via weak-boson fusion at a hadron collider in the form of a flexible parton-level Monte Carlo program which allows us to study cross sections and distributions within experimentally relevant selection cuts. The radiative corrections are found to be moderate with residual scale uncertainties being considerably improved beyond tree level. 
The sensitivity of the reaction to anomalous $W^+W^-\gamma$ couplings is investigated.
\par
\vskip 1cm
\par
\null
\setcounter{page}{0}
\clearpage
\def\thefootnote{\arabic{footnote}}
\setcounter{footnote}{0}
%
%
\section{Introduction}
After the start-up of the CERN Large Hadron Collider (LHC) weak-boson fusion (WBF) reactions deserve attention as a powerful means for discovering the Higgs boson predicted by the Standard Model (SM) and for a later determination of its properties~\cite{ATLAS,CMS,VBF:H,Rainwater:1997dg,VBF:CP,VBF:C,VBF:Hbb}. Moreover, WBF processes will shed light on the mechanism of electroweak symmetry breaking itself, revealing whether the SM is truly realized in nature or extensions are necessary. 
No matter which scenario proves suitable for describing the physics of elementary particles, WBF reactions are expected to provide valuable information on the interactions among weak gauge bosons. Because of the striking kinematic features of this class of processes with well separated tagging jets in the forward and backward regions of the detector, a priori overwhelming QCD backgrounds can be removed efficiently from the electroweak signal signature~\cite{VBF:H,Klamke:2007cu,Rainwater:1997dg}. Radiative corrections have been shown to be well under control~\cite{Figy:2003nv,Berger:2004pca,Ciccolini:2007ec,Oleari:2003tc,Jager:2006zc,Bozzi:2007ur}, and the interference of the purely electroweak WBF processes with QCD production modes giving rise to the same final-state configurations has been found to be negligible~\cite{Ciccolini:2007ec,as:int,Bredenstein:2008tm}. Various WBF processes have thus also been considered as possible means for accessing couplings of the Higgs boson to weak gauge bosons~\cite{VBF:C}, triple and quartic weak boson vertices~\cite{Eboli:2004gc,Eboli:2003nq}, both in the context of the SM and possible extensions thereof. 

In this work, we consider the electroweak (EW) production mode $pp\to \ajj$ at order $\mathcal{O}(\alpha^3)$, which provides access to the coupling of the photon to weak gauge bosons and may thus yield information on this triple gauge boson vertex complementary to bounds derived mainly from single- and double gauge boson production at LEP and the Tevatron~\cite{TGC:EX}. 
We develop a fully flexible parton-level Monte Carlo program, structured similarly to an existing code~\cite{Arnold:2008rz} for related WBF reactions. The program allows for the calculation of cross sections and distributions within experimentally feasible selection cuts. 
Interference effects that are entirely negligible in the phase-space regions where WBF can be observed experimentally~\cite{Ciccolini:2007ec,as:int,Bredenstein:2008tm} are disregarded throughout. In this approximation, the next-to-leading order (NLO) QCD corrections to $pp\to \ajj$ are computed. In order to obtain well-defined predictions for the photon$+2$~jet final state, we employ the isolation criterion of Frixione~\cite{Frixione:1998jh}. 
In addition to discussing the SM production process and estimating the theoretical stability of the prediction, we investigate the sensitivity of the reaction to the $W^+W^-\gamma$ vertex. Although our discussion focusses on $\ajj$ production at the LHC, our results can readily be applied to 
any other high-energy hadron-hadron collider such as the Tevatron. 

We start with a brief outline of the calculational techniques used for developing a parton-level Monte Carlo program in the context of the SM and a summary of the consistency checks we have performed on the thus obtained code in Sec.~\ref{sec:calc}. Numerical results are presented and discussed in Sec.~\ref{sec:num}. Our conclusions are given in Sec.~\ref{sec:conc}. 
%
%
\section{Framework of the calculation}
\label{sec:calc}
\begin{figure}
\bec
\includegraphics[width=0.9\textwidth,clip]{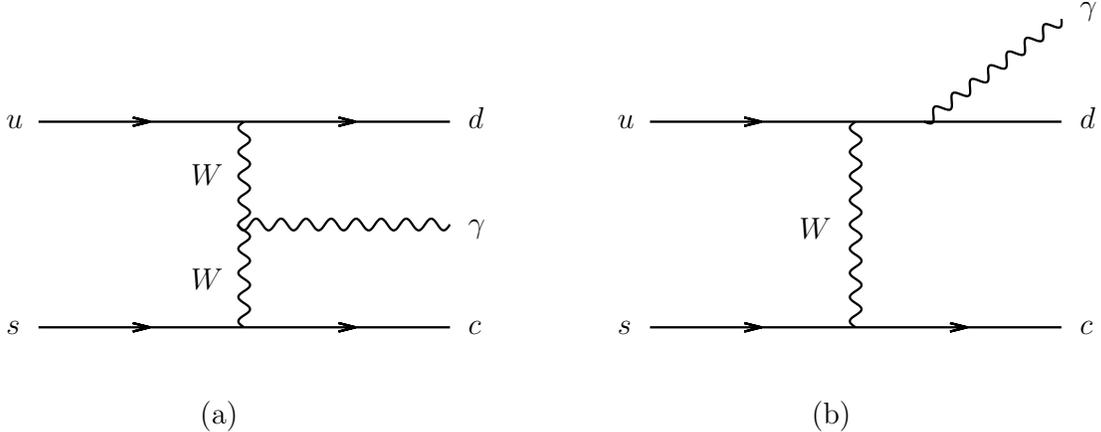}
\caption{
Feynman-diagram topologies contributing to the Born process $us\to dc\gamma$. Graphs analogous to (b), with the photon being emitted off the lower quark line,  are not shown.}
\label{fig:born}
\eec
\end{figure}

At tree-level, EW $\ajj$ production in hadronic collisions mainly proceeds via quark scattering, $qq'\to\gamma qq'$, mediated by the exchange of a weak gauge boson or a photon. 
The relevant charged-current (CC) Feynman-diagrams can be classified in terms of the two topologies depicted in Fig.~\ref{fig:born} for a specific subprocess, depending on whether the photon is being emitted off the $t$-channel exchange boson~(a) or off a fermion line~(b). To neutral-current (NC) $\ajj$ production, only graphs of the latter topology contribute. 
%
%
%
%
Diagrams for the related anti-quark initiated subprocesses $q\bar q'\to q\bar q'\gamma $ and $\bar q \bar q'\to \bar q \bar q'\gamma$ are easily obtained thereof via crossing. 
The evaluation of the relevant tree-level contributions is performed by means of  the amplitude techniques of Ref.~\cite{Hagiwara:1985yu}. 
In the phase-space regions where WBF can be observed experimentally, with widely separated jets of large invariant mass, contributions from annihilation processes such as $q\bar q'\to W^\star\to W\gamma$, with subsequent decay $W\to q\bar q'$, are negligible. Identical-flavor interference effects, which require the interference of $t$-channel with $u$-channel diagrams, are strongly suppressed both at tree-level and beyond in WBF-type reactions, as explicit calculations for the Higgs-production mode have revealed \cite{Ciccolini:2007ec,Bredenstein:2008tm}. We therefore entirely disregard these types of contributions. In the following, we will refer to $pp\to\ajj$ within the mentioned approximations as ``EW $\ajj$ production''.  

The structure of the NLO QCD corrections to EW $\ajj$ production is very similar as in the case of related WBF processes. We therefore do not present details of the calculation reminiscent of what has already been described in the literature \cite{Figy:2003nv,Oleari:2003tc,Jager:2006zc}, but focus on aspects specific to the $\ajj$ final state. 
The real-emission contributions to the CC and NC production modes discussed above are obtained by attaching a gluon to the (anti-)quark lines in all possible ways. Crossing these diagrams yields contributions with an initial-state gluon and an extra anti-quark in the final state, such as $gq'\to \gamma q\bar q q'$.  
Singularities in the soft and collinear regions of phase space are regularized in the dimensional-reduction scheme with space-time dimension $d=4-2 \eps$. The cancellation of these divergences with the respective poles of the virtual contributions is performed by introducing the appropriate counter-terms of the dipole-subtraction method \cite{Catani:1996vz}. Since the color structure of the reaction under consideration is identical to the related  EW $Zjj$ process, the respective counter terms can be straightforwardly adapted from Ref.~\cite{Oleari:2003tc}. 

The virtual contributions are obtained from the interference of the sum of all  one-loop diagrams, $\MV$, with the Born amplitude, $\MB$. Within the approximations outlined above only selfenergy, vertex, and box corrections to either the upper or the lower quark line have to be considered, while contributions from diagrams where a gluon is exchanged between the two fermion lines vanish, due to color conservation. In order to compute the relevant diagrams, we have separately calculated all one-loop corrections to a quark line with only one $t$-channel exchange boson $(V=W^\pm,Z,\gamma)$ being attached, 
\beq
q(k_1)\to V(q_1)\,q(k_2)\,,
\eeq
and to a quark line with the $t$-channel exchange boson and the external photon being attached, 
\beq
\label{eq:qVAq}
q(k_1)\to V(q_1)\,\gamma(q_2)\,q(k_2)\,.
\eeq
The calculation of the virtual contributions is performed analytically by means of {\tt Mathematica}, making use of the {\tt tracer} package \cite{Jamin:1991dp}. 

The interference of all diagrams containing one-loop corrections to either the upper or the lower quark line, $\MV^{(i)}$, with the entire Born amplitude is of the form
\bea
\label{eq:virtual_born}
2\,\mr{Re} \lq \MV^{(i)}\MB^* \rq
&=& |\MB|^2 \frac{\alpha_s(\mur)}{2\pi} C_F
\(\frac{4\pi\mur^2}{Q_i^2}\)^\epsilon \Gamma(1+\epsilon)\\ \non
 &&\times
\lq-\frac{2}{\epsilon^2}-\frac{3}{\epsilon}+c_{\rm virt}\rq\
+2\,\mr{Re} \lq \widetilde{\MV}^{(i)}\MB^* \rq \,,
\eea
where  $Q_i$ is related to the momentum transfer between the respective initial- and the final-state quarks via $Q_i^2 = -(k_1-k_2)^2$, $\mu_R$ is the renormalization scale, $C_F=4/3$, 
$c_{\rm virt}=\pi^2/3-7$, and $\widetilde{\MV}^{(i)}$ is a finite remainder. 

The poles in Eq.~(\ref{eq:virtual_born}) are canceled by respective singularities in the phase-space integrated counter terms, which in the notation of Ref.~\cite{Catani:1996vz} are given by
\beq
\label{eq:I}
\langle \mc{I}(\eps)\rangle  = |\MB|^2 \frac{\alpha_s(\mur)}{2\pi} C_F
\left(\frac{4\pi\mur^2}{Q_i^2}\right)^\epsilon \Gamma(1+\epsilon)
\lq\frac{2}{\epsilon^2}+\frac{3}{\epsilon}+9-\frac{4}{3}\pi^2\rq\;.
\eeq
The non-factorizable terms $\widetilde{\MV}^{(i)}$ can be expressed in terms of the finite parts of the Passarino-Veltman $B_0$, $C_{jk}$, and $D_{jk}$ functions, which are evaluated numerically. 

In order to ensure the reliability of our calculation, several checks have been performed:
\begin{itemize}
\item
The tree-level and real-emission amplitudes have been compared to the corresponding expressions generated automatically by {\tt MadGraph}~\cite{Stelzer:1994ta} for a representative set of phase-space points. We found full agreement. 
\item
The integrated LO cross sections have been compared to the corresponding results of the {\tt MadEvent} package~\cite{Maltoni:2002qb} 
for inclusive selection cuts. 
The cross sections agree within the numerical accuracy of the two programs. 
\item
To validate our implementation of the dipole subtraction, we checked that the real-emission contributions approach the subtraction terms in singular regions of phase space. 
\item
The real-emission matrix elements have been found to vanish when the polarization vector of the external gluon is replaced with its momentum. This procedure tests the QCD gauge invariance of the real-emission amplitudes.
\item
The tree-level, real-emission, and virtual matrix elements have been found to vanish when the polarization vector of the external photon is replaced with its momentum. This procedure tests the QED gauge invariance of the amplitudes.
\item
In order to test our virtual corrections at an analytical level, we have expanded our {\tt Mathematica} modules for the $q\to V\gamma q$ building blocks of Eq.~(\ref{eq:qVAq}) to allow for the simultaneous evaluation of $q\to VZq$.  The latter process is related to $q\bar q\to VZ$ via crossing. Comparing our thus obtained virtual corrections for the $q\bar q\to ZZ$ mode to the literature \cite{Mele:1990bq} at an analytical level provides a valuable check of our calculation. 
\item
In complete analogy to the $\ajj$ case, we have determined the real and virtual NLO-QCD corrections for EW $Zjj$ production, implemented them in our Monte-Carlo program and compared to the code of Ref.~\cite{Oleari:2003tc}. Again, we found full agreement within the numerical accuracy of the two codes, both at amplitude level for selected phase-space points and for integrated cross sections. 
\end{itemize}
%
%
\section{Numerical results and discussion}
\label{sec:num}
The cross-section contributions discussed above have been implemented in a fully flexible parton-level Monte-Carlo program which allows us to compute cross sections and distributions within experimentally relevant selection cuts. 

For our numerical analysis we employ the parton distribution functions of the CTEQ~collaboration with parameterizations being provided both at LO and NLO~QCD. Specifically, we use the CTEQ6M set with $\alpha_s(m_Z) = 0.118$ at NLO, and the CTEQ6L parameterization at LO \cite{Pumplin:2002vw}. For reference, we have also calculated cross sections with the MSTW parton distributions~\cite{Martin:2009iq}, which are available only at NLO~QCD. We found that the respective results differ from those obtained with the CTEQ6~distributions by less than 2\%.
Since in our calculation quark masses are neglected, we entirely disregard contributions from external $b$ and $t$ quarks. As electroweak input parameters, we choose the weak boson masses, $m_Z=91.188$~GeV, $m_W=80.423$~GeV, and the Fermi constant, $G_F=1.166\times 10^{-5}/\mr{GeV}^2$. The other parameters, $\alpha_\mr{QED}$, and $\sin^2\theta_W$, are computed thereof via tree-level electroweak relations. Final-state partons are recombined into jets according to the $k_T$ algorithm~\cite{Catani:1992zp} with a resolution parameter $D=0.7$. If not indicated otherwise, results are shown for $pp$ collisions at the LHC design energy of $\sqrt{S}=14$~TeV. 
For the Cabibbo-Kobayashi-Maskawa matrix, $V_\mr{CKM}$, we have used a 
diagonal form, equal to the identity matrix, which is equivalent to employing 
the exact $V_\mr{CKM}$, when the summation over final-state quark flavors is 
performed and when quark masses are neglected, cf.~Ref~\cite{Bozzi:2007ur}.  

For our phenomenological analysis we utilize 
selection cuts motivated by Ref.~\cite{Rainwater:1997dg} to allow for the identification of WBF events with photons in the final state at the LHC. 
We require at least two hard jets with 
\beq
\label{eq:ptjet-cut}
p_{Tj}\geq 20~\mr{GeV}\,,\quad
|y_j|\leq 4.5\,,
\eeq
where $p_{Tj}$ denotes the transverse momentum and $y_j$ the rapidity of a
jet $j$ being reconstructed from massless partons of
pseudo-rapidity $|\eta_j|< 5$. The two jets of highest transverse
momentum are referred to as ``tagging jets''. We impose a large rapidity
separation between the two tagging jets,
\beq
\Delta y_{jj} = |y_{j1}-y_{j2}|>4.4\,,
\eeq
and furthermore demand that they be located in opposite hemispheres of the detector, 
\beq
y_{j1}\times y_{j2}<0\,,
\eeq
with an invariant mass
\beq
\label{eq:mjj-cut}
M_{jj}>600~\mr{GeV}\,.
\eeq
Imposing the jet-selection criteria of Eqs.~(\ref{eq:ptjet-cut})--(\ref{eq:mjj-cut}) and additional cuts on the final-state photon to be discussed below, the LO differential cross section for EW $\ajj$ production is finite. At NLO, initial-state singularities due to collinear $q\to qg$
and $g\to q\bar q$ splittings can arise. These are taken care of by factorizing
them into the respective quark and gluon distribution functions of the proton.
Additional divergences stemming from the $t$-channel exchange of 
low-virtuality photons in real-emission diagrams are avoided by imposing a cut
on the virtuality of the photon, $Q_{\gamma,min}^2=4$~GeV$^2$. 
Events that do not satisfy the $Q_{\gamma,min}^2$ constraint would give rise
to a $q\rightarrow q\gamma$ collinear singularity, which is part of the
QCD~corrections to $p\gamma\to \ajj$ and not taken into account here.
We have checked that the NLO-QCD cross section within typical WBF cuts is 
quite insensitive to the virtuality cutoff, changing by less than a permille when
$Q_{\gamma,min}^2$ is lowered from 4~GeV$^2$ to 0.1~GeV$^2$. 

To ensure that the final-state photon is well-observable in the central-rapidity range, we require its transverse momentum $p_{T\gamma}$ and rapidity $y_\gamma$ to fulfill the following conditions: 
\bea
&
p_{T\gamma}>20~\mr{GeV} \,,\quad
|y_\gamma|\leq 2.5\,, 
&
\\[2mm]
&
\min(y_{j1},y_{j2})+0.7 \leq y_\gamma \leq \max(y_{j1},y_{j2})-0.7\,.
&
\eea
In order to isolate the photon in an infrared-safe way from partons without having to introduce parton-to-photon fragmentation contributions, we apply the criterion suggested in Ref.~\cite{Frixione:1998jh}: 
An event is considered as acceptable only, if the hadronic energy deposited in a cone around the direction of the photon is limited by 
\beq
\label{eq:acone}
\sum_{i, R_{i\gamma}<R}
      p_{Ti} \leq \frac{1-\cos R}{1-\cos\delta_0} p_{T\gamma} 
\qquad
(\forall R\leq \delta_0)\,. 
\eeq
Here, the summation index $i$ runs over all final-state partons found in a cone of size $R$ in the rapidity-azimuthal angle plane around the photon, $p_{Ti}$ denotes the transverse momentum, and $R_{i\gamma}$ the separation of parton $i$ from the photon, while $\delta_0$ stands for a fixed separation. 
To illustrate the isolation-cone dependence of the integrated cross section for EW~$\ajj$ production within the cuts of Eqs.~(\ref{eq:ptjet-cut})--(\ref{eq:acone}), we list $\sigma^\mr{cuts}$ for various values of $\delta_0$ in Table~\ref{tab:cone-dep}. 
%
%
\begin{table}[!ht]
\begin{center}
\begin{tabular}{|c|c|c|c|c|}
\hline
$\delta_0$ 	 &
$\sigma^\mr{cuts}_\mr{LO}$~[fb]  &
$\sigma^\mr{cuts}_\mr{NLO}$~[fb]   \\
	 \hline
0.50 	 &   4292 	 & 4862	 \\
	 \hline
0.75  	 &   4289  	 & 4827
	\\
	\hline
1.00  	 &   4228	 & 4732
	\\
	\hline
1.25  	 &   4128 	 & 4613
	\\
	\hline
	\hline
\end{tabular}
\caption{
\label{tab:cone-dep}
EW $pp\to\ajj$ cross sections  at
the LHC with $\sqrt{S}=14$~TeV at 
LO and NLO within the cuts of Eqs.~(\ref{eq:ptjet-cut})--(\ref{eq:acone}) for $\mu_F=\mu_R=Q_i$ and four different values of $\delta_0$. The relative statistical errors of the quoted results are at the sub-permille level. 
}	
\vspace*{-.5cm}
\end{center}
\end{table}
%
%
At LO with only two partons in the final state which, due to the selection cuts we impose, are typically well-separated from the photon, the $\delta_0$ dependence is very small, amounting to less than 4\% in the considered range. A slightly larger isolation-cone dependence occurs in the NLO cross sections, as the additional final-state parton in the real-emission contributions can come close to the photon and therefore is more sensitive to the isolation criterion. 
Alternatively to the photon isolation procedure of Ref.~\cite{Frixione:1998jh}, we could opt for computing the full photonic final state, including contributions which stem from parton-to-photon fragmentation. This procedure would allow us to drop the criterion of Eq.~({\ref{eq:acone}), but at the same time introduce a dependence on the so far only poorly known photon fragmentation functions \cite{Buskulic:1995au}. In the following we therefore restrict ourselves to the direct photon production contributions, removing all collinear fragmentation contributions with the help of the isolation procedure described above, and set $\delta_0$ equal to 1.0. 

To illustrate the impact of Eq.~(\ref{eq:acone}) on $\sigma^\mr{cuts}$, in  Fig.~\ref{fig:raj} 
%
%
\begin{figure}[!tp] 
\begin{center}
\includegraphics[width=0.45\textwidth,clip]{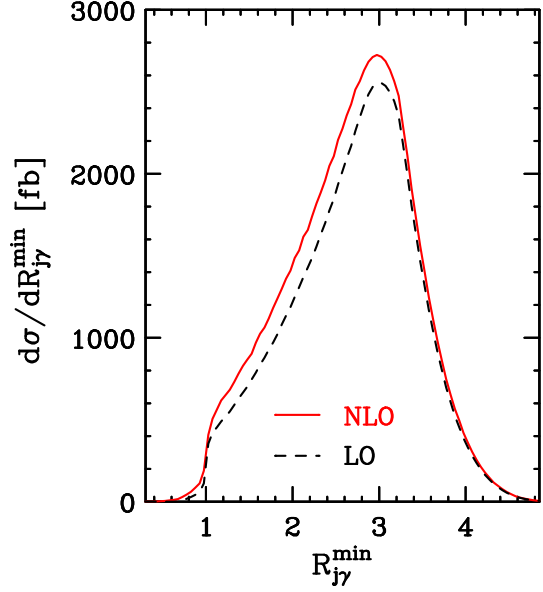}
\end{center}
\vs{-0.5cm}
\caption
{\label{fig:raj} 
Rapidity-azimuthal angle separation of the photon from the closest jet in EW $pp\to \ajj$ production at the LHC with $\sqrt{S}=14$~TeV at 
LO (dashed black line) and NLO (solid red line). }
\end{figure} 
%
%
we show the rapidity-azimuthal angle separation of the photon from the closest jet, $d\sigma/dR_{j\gamma}^\mr{min}$, at LO and NLO. Events with a photon-jet separation smaller than $\delta_0$ yield non-vanishing contributions, unlike what would be observed in the case of an explicit cut on the observable, such as $ R_{j\gamma}>\delta_0$. We furthermore observe that, in complete analogy to the case of EW $Zjj$ production discussed in Ref.~\cite{Oleari:2003tc}, the real-emission contributions shift $d\sigma/dR_{j\gamma}^\mr{min}$ to smaller values, thereby reducing the isolation of the photon from the jets. 

In order to assess the dependence of $\sigma^\mr{cuts}$ on unphysical scales,  in Fig.~\ref{fig:scale-dep} 
%
%
\begin{figure}[!tp] 
\begin{center}
\includegraphics[width=0.45\textwidth,clip]{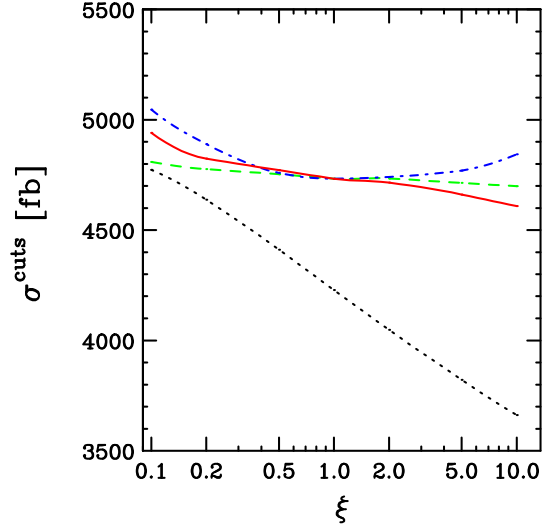}
\end{center}
\vs{-0.5cm}
\caption
{\label{fig:scale-dep} 
Dependence of the EW $pp\to \ajj$  cross section at
the LHC with $\sqrt{S}=14$~TeV on the factorization and renormalization scales. 
The NLO curves show $\sigma^\mr{cuts}$ as a function of the scale parameter
$\xi$ for three different cases: $\mu_R=\mu_F=\xi Q_i$ (solid red line),
$\mu_F=\xi Q_i$ and $\mu_R=Q_i$ (dot-dashed blue line), $\mu_R=\xi Q_i$ and
$\mu_F=Q_i$ (dashed green line). The LO cross sections depend only on $\mu_F$
(dotted black line). 
}
\end{figure} 
%
%
we display the integrated cross section as a function of the renormalization and factorization scales, $\mur$ and $\muf$, which are taken as multiples of the scale parameter $\mu_0$, 
\beq
\muf = \xif\,\mu_0\,,\qquad 
\mur = \xir\,\mu_0\,,
\eeq
where for each fermion line $\mu_0$ is identified with the respective momentum transfer $Q_i$ between the incoming and the outgoing parton. The LO cross section only depends on $\muf$. For the central scale, $\muf=\mur=Q_i$, the NLO-QCD corrections amount to 12\%, being slightly larger than in the case of massive gauge-boson production in WBF-type reactions~\cite{Oleari:2003tc}. 
Qualitatively similar results are obtained for alternative choices of the factorization and renormalization scales such as the average transverse momentum of the $N$ jets in an event, $\sum p_{Tj}/N$. 
In the following, we set $\muf=\mur=Q_i$, unless stated otherwise. 

In addition to the normalization of the integrated cross section, radiative  corrections can affect the shape of various kinematic distributions. To quantify the relative size of NLO-QCD corrections to an observable together with the  residual scale uncertainties, we introduce the quantity $\delta(\mc{O})$, defined by 
\beq
\delta(\mc{O}) = \frac{d\sigma(\muf,\mur)/d\mc{O}}{d\sigma^\mr{NLO}(\muf=\mur=Q_i)/d\mc{O}} - 1\,,
\label{eq:delta}
\eeq
as a measure for the deviation of the LO or NLO expression $d\sigma(\muf,\mur)$, evaluated at an arbitrary scale, from the corresponding NLO observable at the default scale $\muf=\mur=Q_i$.

Figure~\ref{fig:pt-tag}
%
%
%
\begin{figure}[!tp] 
\begin{center}
\includegraphics[width=0.95\textwidth,clip]{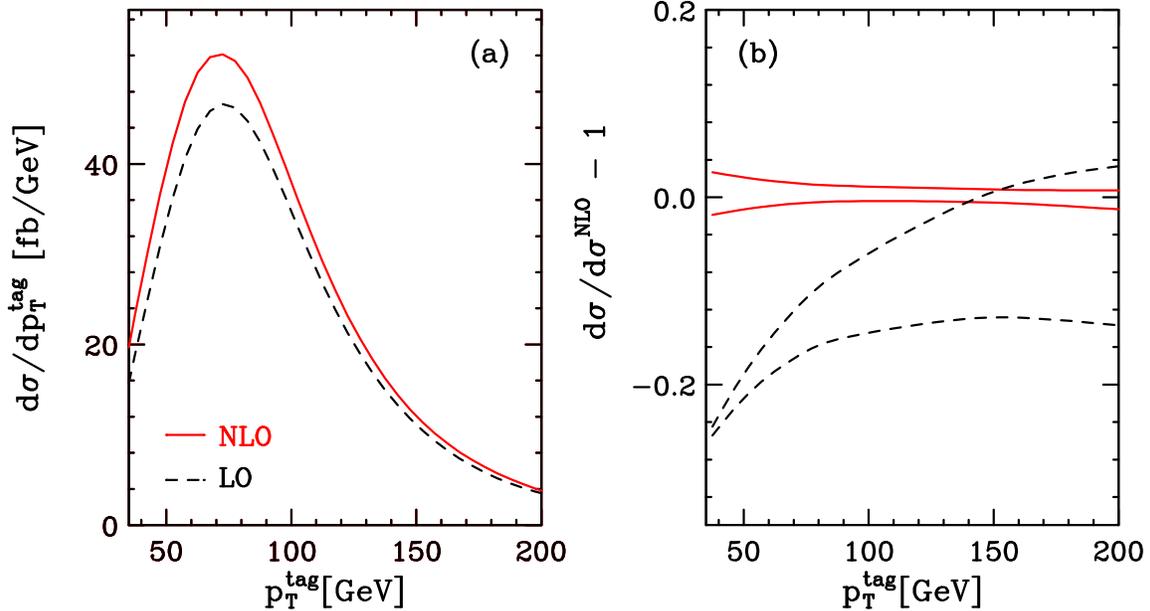}
\end{center}
\vs{-0.5cm}
\caption
{\label{fig:pt-tag} 
Transverse momentum distribution of the highest-$p_T$ tagging jet in EW $pp\to \ajj$ production at the LHC with $\sqrt{S}=14$~TeV at 
LO (dashed black line) and NLO (solid red line) [panel~(a)] and relative corrections according to  Eq.~(\ref{eq:delta}) when the factorization and renormalization scales are varied in the range $Q_i/2\leq\mur=\muf\leq 2Q_i$ [panel~(b)]. }
\end{figure} 
%
%
shows the transverse momentum distribution of the highest-$p_T$ tagging jet at LO and NLO for our default settings, together with the associated relative corrections for two different values of the scale parameter $\xi=\xir=\xif= 1/2$ and $2$.  The difference between the curves for the two values of $\xi$ indicates the scale uncertainty of $d\sigma/dp_{Tj}$ at LO~(dashed black lines) and NLO~(solid red lines), respectively. 
At low transverse momenta, relatively large positive corrections occur. 
Towards higher values of $p_{Tj}$ the impact of NLO-QCD corrections on the cross section decreases. In this kinematic range the scale uncertainty of the LO distribution is sizeable, but can be efficiently reduced by taking the NLO-QCD contributions into account. Similar features are observed for lower values of the hadronic center-of-mass~energy, cf.~Fig.~\ref{fig:pt-tag-7t}. 
%
%
\begin{figure}[!tp] 
\begin{center}
\includegraphics[width=0.95\textwidth,clip]{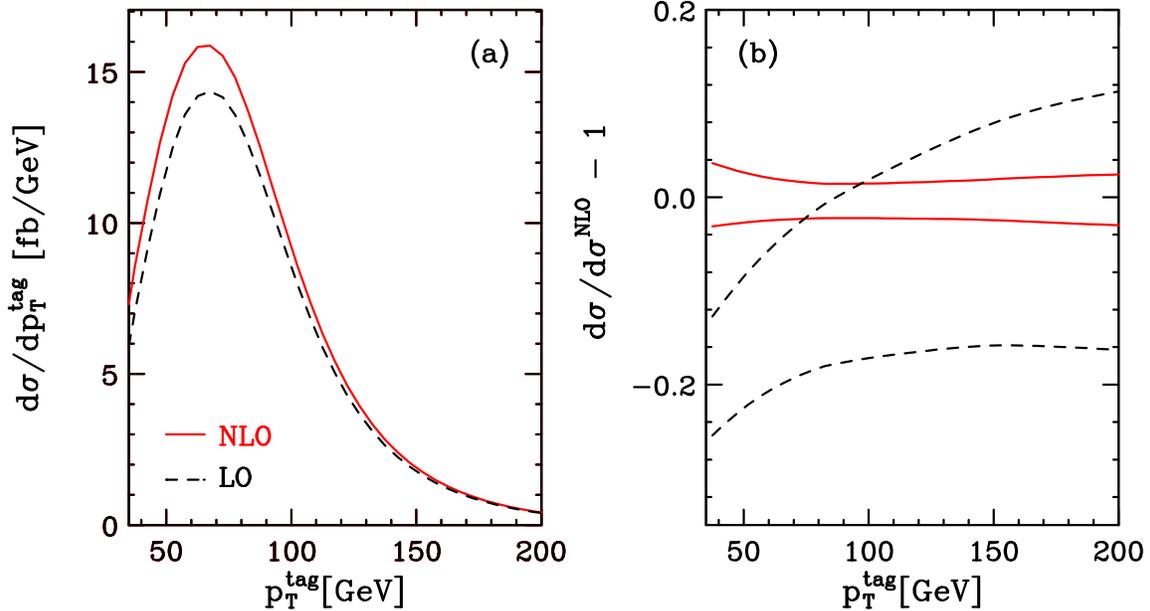}
\end{center}
\vs{-0.5cm}
\caption
{\label{fig:pt-tag-7t} 
Transverse momentum distribution of the highest-$p_T$ tagging jet in EW $pp\to \ajj$ production at the LHC with $\sqrt{S}=7$~TeV at 
LO (dashed black line) and NLO (solid red line) [panel~(a)] and relative corrections according to  Eq.~(\ref{eq:delta}) when the factorization and renormalization scales are varied in the range $Q_i/2\leq\mur=\muf\leq 2Q_i$ [panel~(b)]. }
\end{figure} 
%
%
Decreasing $\sqrt{S}$ from 14~TeV to 7~TeV yields smaller cross sections and slightly larger scale uncertainties with radiative corrections again affecting predominantly the low-$p_{Tj}$ region. 

Numerically, the EW $\ajj$ cross section is dominated by CC~subprocesses, where the two incoming partons scatter by $t$-channel $W$-boson exchange. It is thus particularly suitable to access the $W^+W^-\gamma$ coupling. 
Since NLO-QCD corrections do not affect the weak boson vertex, the implementation of anomalous triple gauge boson couplings~(TGC) in our Monte-Carlo program is straightforward, allowing us to study their impact on the cross section and kinematic distributions. 
In the effective Lagrangian approach of Ref.~\cite{Hagiwara:1986vm}, the Lorentz structure of the momentum-space vertex $W^-_\alpha(q) W^+_\beta(q') \gamma_\mu(p)$ (with all momenta outgoing, $p + q + q' = 0$) is parameterized in the form 
\bea
 \Gamma^{\alpha \beta \mu}_{W W \gamma}(q, q', p) &=& 
   q'^\alpha g^{\beta \mu} 
    \biggl( 2 + \Delta\kappa^\gamma + \lambda^\gamma {q^2\over m_W^2} \biggr) 
 - q^\beta g^{\alpha \mu}
    \biggl( 2 + \Delta\kappa^\gamma + \lambda^\gamma {q'^2\over m_W^2}
\biggr) \nn \\  
&& 
 + \bigl( q'^\mu - q^\mu \bigr) 
 \Biggl[ - g^{\alpha \beta} \biggl( 
   1 + {1\over2} p^2 \frac{\lambda^\gamma}{m_W^2} \biggr) 
 +\frac{\lambda^\gamma}{m_W^2} p^\alpha p^\beta \Biggr] \,, 
\label{eq:WWa-vertex}
\eea
where $C$ and $P$ invariance have been imposed and only operators up to dimension six are taken into account. For an on-shell photon we have $p^2=0$, and thus the third term in the above expression further simplifies. In the limit $\Delta\kappa^\gamma\to 0$, $\lambda^\gamma\to 0$, Eq.~(\ref{eq:WWa-vertex}) reduces to the SM expression for the $W^+W^-\gamma$ vertex, which is compatible with combined limits from LEP and the Tevatron~\cite{Amsler:2008zzb}. 
A priori, the effective Lagrangian approach gives rise to unitarity violations in the high-energy regime. To avoid this unphysical behavior, the anomalous gauge boson couplings in Eq.~(\ref{eq:WWa-vertex}) have to be supplemented by a form factor. Following the prescription of Ref.~\cite{Eboli:2004gc}, we perform the substitutions  
\bea
\Delta\kappa^\gamma \to \frac{\Delta\kappa^\gamma}{
\parbox{24ex}{\vs{.1cm}$
\Bigl[\Bigl(1+\frac{|q^2|}{\Lambda^2}\Bigr)
      \Bigl(1+\frac{|q'^2|}{\Lambda^2}\Bigr)\Bigr]^n$}
      }\,,
\qquad
\lambda^\gamma \to \frac{\lambda^\gamma}{
\parbox{26ex}{\vs{.1cm}$
\Bigl[\Bigl((1+\frac{|q^2|}{\Lambda^2}\Bigr)
      \Bigl((1+\frac{|q'^2|}{\Lambda^2}\Bigr)\Bigr]^n$}
      }\,,
\label{eq:formfac}
\eea
where $n$ is an exponent that has to be chosen large enough to ensure unitarity and $\Lambda$ is
interpreted as the scale where new physics enters that is not accounted for by the effective Lagrangian approach. Below, we will set $\Lambda = 2$~TeV and $n=1$. We will show, however, that the impact of the form factor on the observables we consider is rather small. 

Anomalous gauge boson couplings mainly affect events at high energy or large transverse momenta. Therefore, the high-$p_{T\gamma}$ tail of the photon's transverse momentum distribution is expected to be particularly sensitive to $\Delta\kappa^\gamma$ and $\lambda^\gamma$. 
For being able to distinguish the anomalous coupling effects from radiative corrections, a precise knowledge of this distribution both within the SM and its considered extension is essential.  
The SM prediction, $d\sigma/dp_{T\gamma}$, is displayed at LO and NLO together with the associated relative corrections for different values of the factorization and renormalization scales in Fig.~\ref{fig:pta}.  
%
%
%
\begin{figure}[!tp] 
\begin{center}
\includegraphics[width=0.95\textwidth,clip]{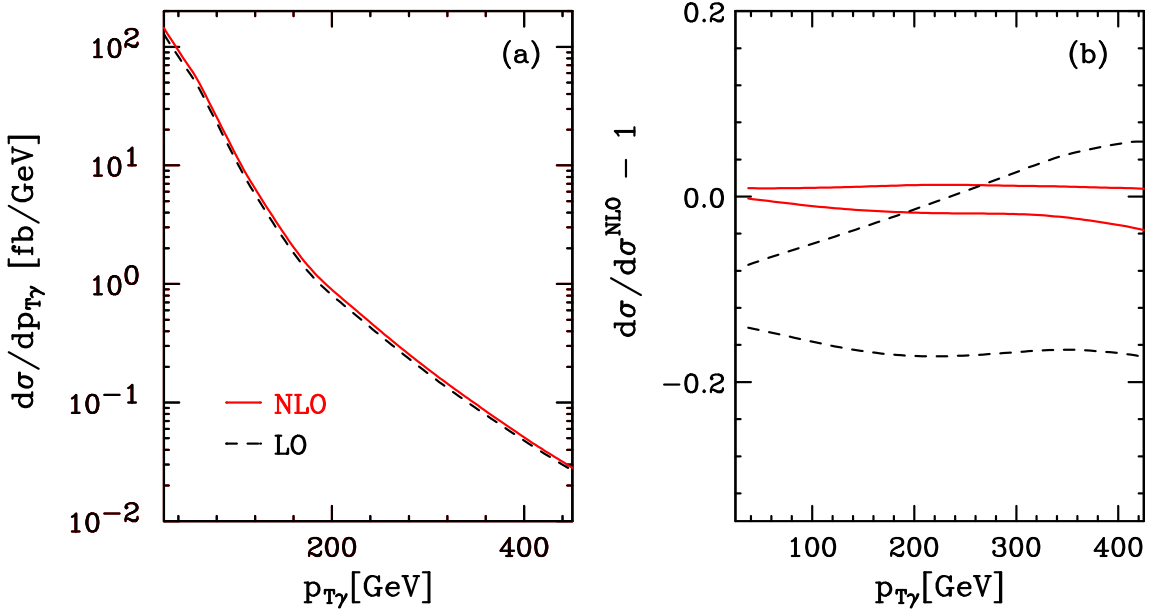}
\end{center}
\vs{-0.5cm}
\caption
{\label{fig:pta} 
Transverse momentum distribution of the photon in EW $pp\to \ajj$ production at the LHC with $\sqrt{S}=14$~TeV at 
LO (dashed black line) and NLO (solid red line) [panel~(a)] and relative corrections according to  Eq.~(\ref{eq:delta}) when the factorization and renormalization scales are varied in the range $Q_i/2\leq\mur=\muf\leq 2Q_i$ [panel~(b)]. 
}
\end{figure} 
%
%
The NLO-QCD corrections to this distribution are moderate, but distort its shape noticeably. Reminiscent of our observations for the transverse momentum distribution of the high-$p_T$ tagging jet, we find that in the low transverse momentum range relatively large positive corrections occur. The scale dependence increases with $p_{T\gamma}$, but is considerably improved when going from LO to NLO. Having estimated the uncertainties of the SM distribution, we now turn to an analysis of the impact anomalous TGC may have on $d\sigma/dp_{T\gamma}$. 

To this end, we consider the ratio of the distribution with non-vanishing anomalous TGC, $d\sigma^\mr{NLO}_\mr{AC}/dp_{T\gamma}$, to the respective SM curve, $d\sigma^\mr{NLO}_\mr{SM}/dp_{T\gamma}$, depicted in 
Fig.~\ref{fig:pta-anom} 
%
%
\begin{figure}[!tp] 
\begin{center}
\includegraphics[width=0.95\textwidth,clip]{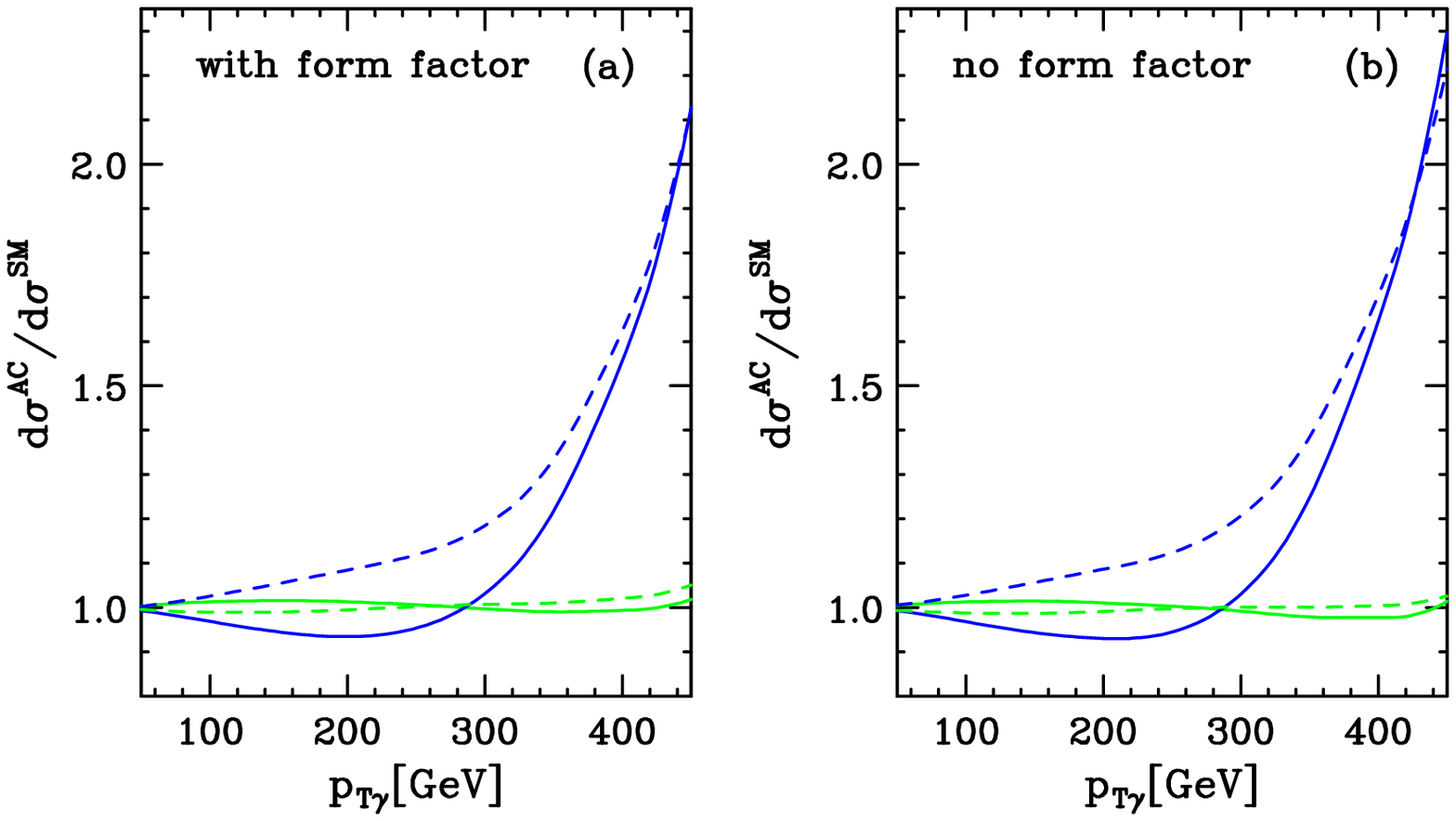}
\end{center}
\vs{-0.5cm}
\caption
{\label{fig:pta-anom} 
Ratio of $d\sigma^\mr{NLO}_\mr{AC}/dp_{T\gamma}$ and $d\sigma^\mr{NLO}_\mr{SM}/dp_{T\gamma}$ in EW $pp\to\ajj$ production at the LHC with $\sqrt{S}=14$~TeV for 
$\Delta\kappa^\gamma= 0.02$, $\lambda^\gamma=0$ (dashed green lines), for $\Delta\kappa^\gamma=-0.02$, $\lambda^\gamma=0$ (solid green lines), for $\Delta\kappa^\gamma=0$, $\lambda^\gamma= 0.02$ (dashed blue lines), and for $\Delta\kappa^\gamma=0$, $\lambda^\gamma=-0.02$ (solid blue lines). In panel~(a) the form factor of Eq.~(\ref{eq:formfac}) is included, while in panel~(b) no form factor has been supplemented. 
}
\end{figure} 
%
%
for two scenarios: In panel~(a) the form factor of Eq.~(\ref{eq:formfac}) is included, while in panel~(b) no form factor has been supplemented. 
The two parameters $\Delta\kappa^\gamma$ and $\lambda^\gamma$ are set to non-zero values compatible with current experimental limits. 
We find that $d\sigma/dp_{T\gamma}$  is sensitive to $\lambda^\gamma$ over the entire transverse momentum range. As expected, a non-zero value of $\lambda^\gamma$  becomes particularly evident towards the high-$p_{T\gamma}$~tail. 
The dependence on $\Delta\kappa^\gamma$ is less pronounced. The impact of the form factor is small, as becomes evident from contrasting the results in panel~(a) with the respective curves in panel~(b). Only at very large values of $p_{T\gamma}$ the distributions without a form factor start to exceed the corresponding predictions including a form factor. 
We note that anomalous TGC give rise to similar shape distortions in the tagging jets' transverse momentum distributions. 

Further information on the structure of the $W^+W^-\gamma$ vertex can be obtained from the azimuthal-angle separation of the two tagging jets, $\Delta\phi_{jj}$. Azimuthal-angle correlations have been suggested as a tool for determining the $CP$ properties of the Higgs boson in $pp\to~Hjj$~\cite{VBF:CP}. Eboli and Garcia~\cite{Eboli:2004gc} have furthermore considered $d\sigma/d\Delta\phi_{jj}$ as a means for constraining anomalous TGC in $pp\to W^\pm jj$. 
Figure~\ref{fig:phijj-anom} 
%
%
\begin{figure}[!tp] 
\begin{center}
\includegraphics[width=0.95\textwidth,clip]{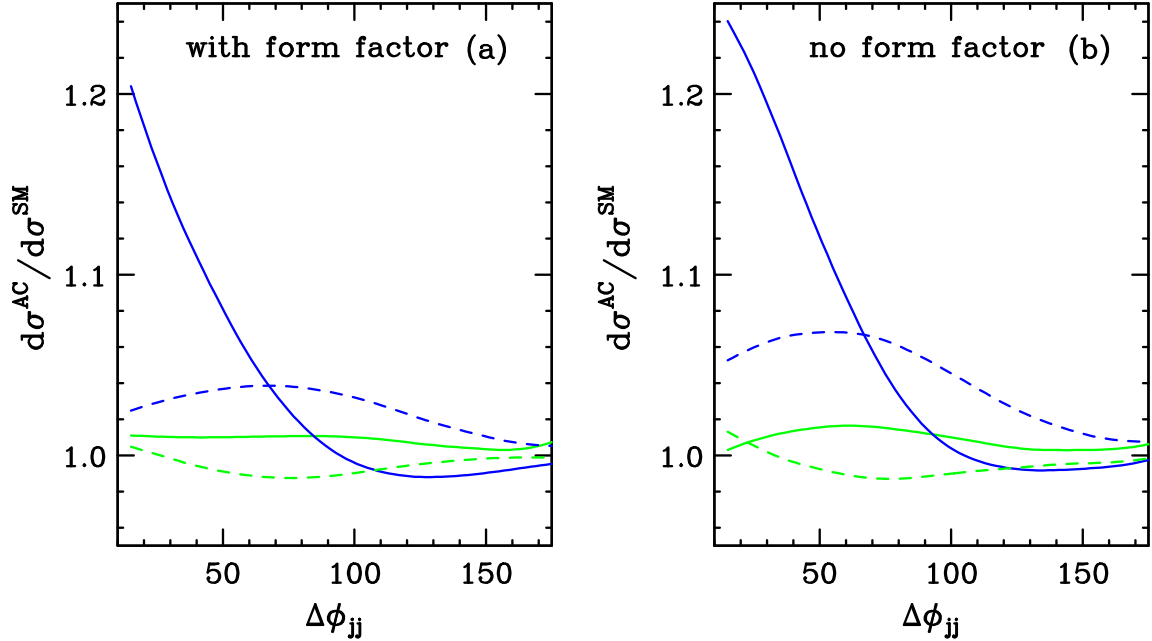}
\end{center}
\vs{-0.5cm}
\caption
{\label{fig:phijj-anom} 
Ratio of $d\sigma^\mr{NLO}_\mr{AC}/d\Delta\phi_{jj}$ and $d\sigma^\mr{NLO}_\mr{SM}/d\Delta\phi_{jj}$  in EW $pp\to\ajj$ production at the LHC with $\sqrt{S}=14$~TeV for 
$\Delta\kappa^\gamma= 0.02$, $\lambda^\gamma=0$ (dashed green lines), for $\Delta\kappa^\gamma=-0.02$, $\lambda^\gamma=0$ (solid green lines), for $\Delta\kappa^\gamma=0$, $\lambda^\gamma= 0.02$ (dashed blue lines), and for $\Delta\kappa^\gamma=0$, $\lambda^\gamma=-0.02$ (solid blue lines). In panel~(a) the form factor of Eq.~(\ref{eq:formfac}) is included, while in panel~(b) no form factor has been supplemented. 
}
\end{figure} 
%
%
shows this distribution, normalized to the SM result, for various values of $\Delta\kappa_\gamma$ and $\lambda_\gamma$ in $pp\to \ajj$. A marked enhancement is observed at low values of $\Delta\phi_{jj}$ for non-zero values of $\lambda_\gamma$, while the impact of $\Delta\kappa^\gamma$ is very small. Including the form factor of Eq.~(\ref{eq:formfac}) does not change the qualitative features of the distribution. 
%
%
\section{Conclusions}
\label{sec:conc}
In this article we have presented results for EW $\ajj$ production in hadronic collisions at NLO-QCD accuracy, obtained with a fully-flexible parton-level Monte Carlo program which allows us to study cross sections and distributions within realistic experimental selection cuts. NLO-QCD corrections were found to enhance integrated cross sections by about 12\% and change the shape of some distributions. The residual scale dependence of all results is significantly improved beyond the Born approximation. After the inclusion of NLO-QCD corrections, the cross section within the WBF-specific selection cuts we have introduced in Sec.~\ref{sec:num} changes by less than 2\% when factorization and renormalization scale are varied simultaneously in the range $Q_i/2\leq \muf=\mur\leq 2Q_i$. Compared to the more pronounced scale dependence of about 9\% at LO, this improvement indicates that the process is under excellent control perturbatively. 

Moreover, we found that $pp\to \ajj$ could yield information on the $W^+W^-\gamma$ coupling complementary to bounds derived from single and double gauge boson production at LEP and the Tevatron. To demonstrate the sensitivity of the EW $\ajj$ cross section on this vertex, we implemented anomalous TGC in our code and demonstrated that 
noticeable changes are induced by non-SM contributions to the $W^+W^-\gamma$ vertex on selected observables, such as the transverse momentum distributions of the photon and of the tagging jets, as well as the azimuthal angle separation of the tagging jets. 
Since anomalous gauge boson couplings affect the shape of these distribution in a manner very different from NLO-QCD contributions, their sensitivity is not spoiled by radiative corrections. 
%
%
\section*{Acknowledgments}
I am grateful to S.~Dittmaier, S.~Frixione, A.~Kulesza, and M.~Stratmann for valuable comments and discussions. This work has been supported by the Initiative and Networking Fund of the Helmholtz Association, contract HA-101 ("Physics at the Terascale"). 
%
%

\end{document}

%% file: setup.tex
\usepackage{graphics}
\usepackage{amsmath}
\usepackage{amssymb}
\usepackage{cite}
\usepackage{eufrak}
\usepackage{rotating}
\usepackage{wasysym}

\makeatletter
\if@twoside \oddsidemargin -17pt \evensidemargin 00pt
\else \oddsidemargin 00pt \evensidemargin 00pt
\fi
\expandafter\ifx\csname mathrm\endcsname\relax\def\mathrm#1{{\rm #1}}\fi

\makeatother

\def\beq{\begin{equation}}
\def\eeq{\end{equation}}
\def\bec{\begin{center}}
\def\eec{\end{center}}
\def\nn{\nonumber}

\newcommand{\bea}{\begin{eqnarray}}
\newcommand{\eea}{\end{eqnarray}}
\newcommand{\non}{\nonumber}
\newcommand{\mc}{\mathcal}

\newcommand{\mr}{\mathrm}

\newcommand{\vs}{\vspace*}

\def\muf{\mu_\mr{F}}
\def\mur{\mu_\mr{R}}
\def\xif{\xi_\mr{F}}
\def\xir{\xi_\mr{R}}
\def\eps{\varepsilon}
\def\ajj{\gamma j j }
\def\MB{{\cal M}_B}
\def\MV{{\cal M}_V}

\def\lq{\left[} 
\def\rq{\right]}

\def\({\left(} 
\def\){\right)} 


%% file: ajj.bbl
\begin{thebibliography}{99}
\frenchspacing
%
\bibitem{ATLAS}
ATLAS Collaboration, ATLAS TDR,
Report No.\ CERN/LHCC/99-15 (1999).

\bibitem{CMS}
G.~L.~Bayatian {\it et al.}, CMS TDR, 
Report No.\ CERN/LHCC/2006-021 (2006).

\bibitem{VBF:H}
D.~Rainwater and D.~Zeppenfeld,
Phys.\ Rev.\ D {\bf 60}, 113004 (1999)
[Erratum-ibid.\ D {\bf 61}, 099901 (2000)]
[arXiv:hep-ph/9906218];
N.~Kauer, T.~Plehn, D.~Rainwater, and D.~Zeppenfeld,
Phys.\ Lett.\ B {\bf 503}, 113 (2001)
[arXiv:hep-ph/0012351].

\bibitem{Rainwater:1997dg}
D.~L.~Rainwater and D.~Zeppenfeld, 
JHEP {\bf 12}, 005 (1997) 
[arXiv:hep-ph/9712271]. 

\bibitem{VBF:CP}
T.\ Plehn, D.\ Rainwater, and D.\ Zeppenfeld, 
Phys.\ Rev.\ Lett.\ {\bf 88}, 051801 (2002) 
[arXiv:hep-ph/0105325];
V.\ Hankele, G.\ Kl{\"a}mke, D.\ Zeppenfeld, and T.\ Figy, 
Phys.\ Rev.\ D {\bf 74}, 095001 (2006)
[arXiv:hep-ph/0609075].
 
\bibitem{VBF:C}
D.~Zeppenfeld, R.~Kinnunen, A.~Nikitenko, and E.~Richter-Was,
Phys.\ Rev.\ D {\bf 62}, 013009 (2000)
[arXiv:hep-ph/0002036];
M.~D\"uhrssen {\it et al.},
Phys.\ Rev.\ D {\bf 70}, 113009 (2004)
[arXiv:hep-ph/0406323].

\bibitem{VBF:Hbb}
E.~Gabrielli {\it et al.}, 
Nucl.~Phys.~B~{\bf 781}, 64 (2007) 
[arXiv:hep-ph/0702119]; 
%
T.~Han and B.~Mellado,
arXiv:0909.2460; 
%
B.~J\"ager, 
Phys.~Rev.~D~{\bf 81}, 054018 (2010) 
[arXiv:1001.3789].

\bibitem{Klamke:2007cu}
G.~Kl\"amke and D.~Zeppenfeld,
JHEP~{\bf 04}, 052 (2007) 
[arXiv:hep-ph/0703202].

\bibitem{Figy:2003nv}
T.~Figy, C.~Oleari, and D.~Zeppenfeld,
Phys.\ Rev.\ D {\bf 68}, 073005 (2003)
[arXiv:hep-ph/0306109].  

\bibitem{Berger:2004pca}
E.\ L.\ Berger and J.\ Campbell,
Phys.\ Rev.\ D {\bf 70}, 073011 (2004) 
[arXiv:hep-ph/0403194].

\bibitem{Ciccolini:2007ec}
M.~Ciccolini, A.~Denner, and S.~Dittmaier,
Phys.\ Rev.\  D {\bf 77}, 013002 (2008)
[arXiv:0710.4749].

\bibitem{Oleari:2003tc}
C.~Oleari and D.~Zeppenfeld, 
Phys.~Rev.~D~{\bf 69}, 093004 (2004) 
[arXiv:hep-ph/0310156]. 

\bibitem{Jager:2006zc} 
B.~J\"ager, C.~Oleari, and D.~Zeppenfeld,
JHEP {\bf 0607}, 015 (2006) 
[arXiv:hep-ph/0603177]; 
%
Phys.\ Rev.\ D {\bf 73}, 113006 (2006)
[arXiv:hep-ph/0604200];
%
Phys.~Rev.~D~{\bf 80}, 034022 (2009) 
[arXiv:0907.0580].

\bibitem{Bozzi:2007ur} 
G.~Bozzi, B.~J\"ager, C.~Oleari, and D.~Zeppenfeld,
Phys.\ Rev.\ D {\bf 75}, 073004 (2007)
[arXiv:hep-ph/0701105].

\bibitem{as:int} 
J.\ R.\ Andersen and J.\ M.\ Smillie, 
Phys.\ Rev.\ D {\bf 75}, 037301 (2007) 
[arXiv:~hep-ph/0611281].

\bibitem{Bredenstein:2008tm}
A.~Bredenstein, K.~Hagiwara, and B.~J\"ager,
Phys.~Rev.~D~{\bf 77}, 073004 (2008)
[arXiv:0801.4231]. 

\bibitem{Eboli:2004gc}
O.~J.~P.~Eboli and M.~C.~Gonzalez-Garcia,  
Phys.~Rev.~D~{\bf 70}, 074011 (2004)
[arXiv:hep-ph/0405269].

\bibitem{Eboli:2003nq}
O.~J.~P.~Eboli, M.~C.~Gonzalez-Garcia, and S.~M.~Lietti,
Phys.~Rev.~D~{\bf 69}, 095005 (2004) 
[arXiv:hep-ph/0310141];
%
O.~J.~P.~Eboli, M.~C.~Gonzalez-Garcia, and J.~K.~Mizukoshi,
Phys.~Rev.~D~{\bf 74},
073005 (2006) 
[arXiv:hep-ph/0606118].

\bibitem{TGC:EX}
ALEPH, DELPHI, L3, and OPAL Collaborations, LEP Electroweak Working Group, and SLD Heavy Flavor Group, 
arXiv:hep-ex/0212036; 
%
B.~Abbott {\it et al.}, 
Phys.~Rev.~D~{\bf 62},
052005 (2000) 
[arXiv:hep-ex/9912033];
%
T.~Aaltonen {\it et al.}, 
Phys.~Rev.~D~{\bf 76}, 
111103 (2007) 
[arXiv:hep-ex/0705.2247].

\bibitem{Arnold:2008rz}
K.~Arnold {\it et al.}, 
Comput.~Phys.~Commun.~{\bf 180}, 1661 (2009) 
[arXiv:0811.4559].

\bibitem{Frixione:1998jh}
S.~Frixione, 
Phys.~Lett.~B~{\bf 429}, 369 (1998)
[arXiv:hep-ph/9801442].

\bibitem{Hagiwara:1985yu}
K.~Hagiwara and D.~Zeppenfeld,
Nucl.\ Phys.\ {\bf B274}, 1 (1986);
%
Nucl.\ Phys.\  {\bf B313}, 560 (1989).

\bibitem{Catani:1996vz}
S.~Catani and M.~H.~Seymour,
Nucl.\ Phys.\  {\bf B485}, 291 (1997)
[Erratum-ibid.\  {\bf B510}, 503 (1997)]
[arXiv:hep-ph/9605323].

\bibitem{Jamin:1991dp}
M.~Jamin and M.~E.~Lautenbacher, 
Comput.~Phys.~Commun.~{\bf 74}, 265 (1993). 

\bibitem{Stelzer:1994ta}
T.~Stelzer and W.~F.~Long,
Comput.\ Phys.\ Commun.\  {\bf 81}, 357 (1994)
[arXiv:hep-ph/9401258]. 

\bibitem{Maltoni:2002qb}
F.~Maltoni and T.~Stelzer,
JHEP {\bf 0302}, 027 (2003)
[arXiv:hep-ph/0208156]; 
%
J.\ Alwall {\it et al.}, 
JHEP {\bf 0709}, 028 (2007)
[arXiv:0706.2334].

\bibitem{Mele:1990bq}
B.~Mele, P.~Nason, and G.~Ridolfi, 
Nucl.~Phys.~{\bf B357}, 409 (1991). 

\bibitem{Pumplin:2002vw} 
J.~Pumplin {\it et al.}, 
JHEP {\bf 0207}, 012 (2002)
[arXiv:hep-ph/0201195].

\bibitem{Martin:2009iq}
A.~D.~Martin, W.~J.~Stirling, R.~S~Thorne, and G.~Watt, 
Eur.~Phys.~J.~C~{\bf 63}, 189 (2009), 
[arXiv:0901.0002]. 

\bibitem{Catani:1992zp}
S.~Catani, Yu.~L.~Dokshitzer, and B.~R.~Webber,
Phys.\ Lett.\ B  {\bf 285}, 291 (1992);  
%
S.~Catani, Yu.~L. Dokshitzer, M.~H.~Seymour, and B.~R.~Webber,
Nucl.\ Phys.\  {\bf B406}, 187 (1993);
%
S.~D.~Ellis and D.~E.~Soper, Phys.\ Rev.\ D
{\bf 48}, 3160 (1993)
[arXiv:hep-ph/9305266];
%
G.~C.~Blazey {\it et al.},
arXiv:hep-ex/0005012.

\bibitem{Buskulic:1995au}
D.~Buskulic {\it et al.}, 
Z.~Phys.~C~{\bf 69}, 365 (1996);
%
K.~Ackerstaff {\it et al.}, 
Eur.~Phys.~J.~C~{\bf 2},
39 (1998) 
[arXiv:hep-ex/9708020];
%
M.~Gl\"uck, E.~Reya, and A.~Vogt, 
Phys.~Rev.~D~{\bf 48}, 
116 (1993);
%
L.~Bourhis, M.~Fontannaz, and J.~P.~Guillet, 
Eur.~Phys.~J.~C~{\bf 2},
529 (1998) 
[arXiv:hep-ph/9704447].

\bibitem{Hagiwara:1986vm}
K.~Hagiwara, R.~D.~Peccei, D.~Zeppenfeld, and K.~Hikasa, 
Nucl.~Phys.~B~{\bf 282}, 
253 (1987).
%
L.~J.~Dixon, Z.~Kunszt, and A.~Signer, 
Phys.~Rev.~D~{\bf 60}, 
114037 (1999) 
[arXiv:hep-ph/9907305].

\bibitem{Amsler:2008zzb}
C.~Amsler {\it et al.}, 
Phys.~Lett.~B~{\bf 667}, 1 (2008). 

\end{thebibliography}
